\documentclass[letter]{aa} 

\usepackage{graphicx}
\usepackage[colorlinks=true,urlcolor=blue,citecolor=blue,linkcolor=blue]{hyperref}
\usepackage{txfonts}

\begin{document} 

\title{Propagation of transverse waves in the solar chromosphere probed at different heights with ALMA sub-bands}

   \author{Juan Camilo Guevara G\'omez \inst{1,2}
          \and
           Shahin Jafarzadeh  \inst{3,1} 
          \and
          Sven Wedemeyer  \inst{1,2}
          \and
          Mikolaj Szydlarski  \inst{1,2}}
   
  \authorrunning{Guevara G\'omez {et~al.}}
  \titlerunning{Transverse waves and height difference in ALMA sub-bands}
  
   \institute{Rosseland Centre for Solar  Physics, University of Oslo, Postboks 1029 Blindern, 0315 Oslo, Norway\\
   \email{j.c.g.gomez@astro.uio.no}
            \and
            Institute of  Theoretical Astrophysics, University of Oslo, Postboks 1029 Blindern, 0315 Oslo, Norway 
            \and
            Max Planck Institute for Solar System Research, Justus-von-Liebig-Weg 3, 37077 G\"{o}ttingen, Germany\\
} 

   \date{Received: 30 June 2022 / Accepted: 20 August 2022}

\abstract{The Atacama Large Millimeter/sub-millimeter Array (ALMA) has provided us with an excellent diagnostic tool for studies of the dynamics of the Solar chromosphere, albeit through a single receiver band at one time presently. Each ALMA band consists of four sub-bands that are comprised of several spectral channels. To date, however, the spectral domain has been neglected in favour of ensuring optimal imaging, so that time-series observations have been mostly limited to full-band data products,  thereby limiting studies to a single chromospheric layer. Here, we report the first observations of a dynamical event (i.e. wave propagation) for which the ALMA Band 3 data (centred at 3\,mm; 100\,GHz) is split into a lower and an upper sideband. In principle, this approach is aimed at mapping slightly different layers in the Solar atmosphere.  The side-band data were reduced together with the Solar ALMA Pipeline (SoAP), resulting in time series of brightness-temperature maps for each side-band. Through a phase analysis of a magnetically quiet region, where purely acoustic waves are expected to dominate, the average height difference between the two side-bands is estimated as $73\pm16$~km. Furthermore, we examined the propagation of transverse waves in small-scale bright structures by means of wavelet phase analysis between oscillations at the two atmospheric heights. We find 6\% of the waves to be standing, while 54\% and 46\% of the remaining waves are propagating upwards and downwards, respectively, with absolute propagating speeds on the order of $\approx96$~km/s, resulting in a mean energy flux of $3800$\,W/m$^2$.}


\maketitle


\section{Introduction}

The Solar chromosphere is a highly dynamic environment where interactions between the magnetic fields and plasma occur across a broad range of spatial and temporal scales \citep{2009ApJ...706..148W, 2019ARA&A..57..189C}. In particular, waves and oscillations play an important role in transferring  energy and momentum throughout the atmosphere, thus maintaining the energy balance of the chromosphere and beyond \citep{1993ApJ...413..811C, 2008ApJ...680.1542H}. While  oscillatory phenomena and their propagation through the Solar chromosphere have readily been studied for more than half a century \citep{2015SSRv..190..103J, 2015LRSP...12....6K}, direct observations of their energy deposition, particularly on small scales, have been challenging \citep{2017ApJS..229....7G, 2017ApJS..229....9J}. This is partly due to the commonly used chromospheric diagnostics being subject to non-local thermodynamic equilibrium (non-LTE) effects, which have made it difficult to reliably infer the physical parameters  \citep{2017SSRv..210..109D}. Alternatively, observations at millimetre wavelengths (which are formed under LTE conditions and are optically thick) would provide direct observations of brightness temperatures, serving as a close proxy for the local electron temperature   \citep{2016SSRv..200....1W, 2017SoPh..292...88W,2019ApJ...881...99M,2020arXiv200512717C,2021A&A...652A..92N}.

\begin{figure*}[tp!]
    \centering
    \includegraphics[width=.95\textwidth]{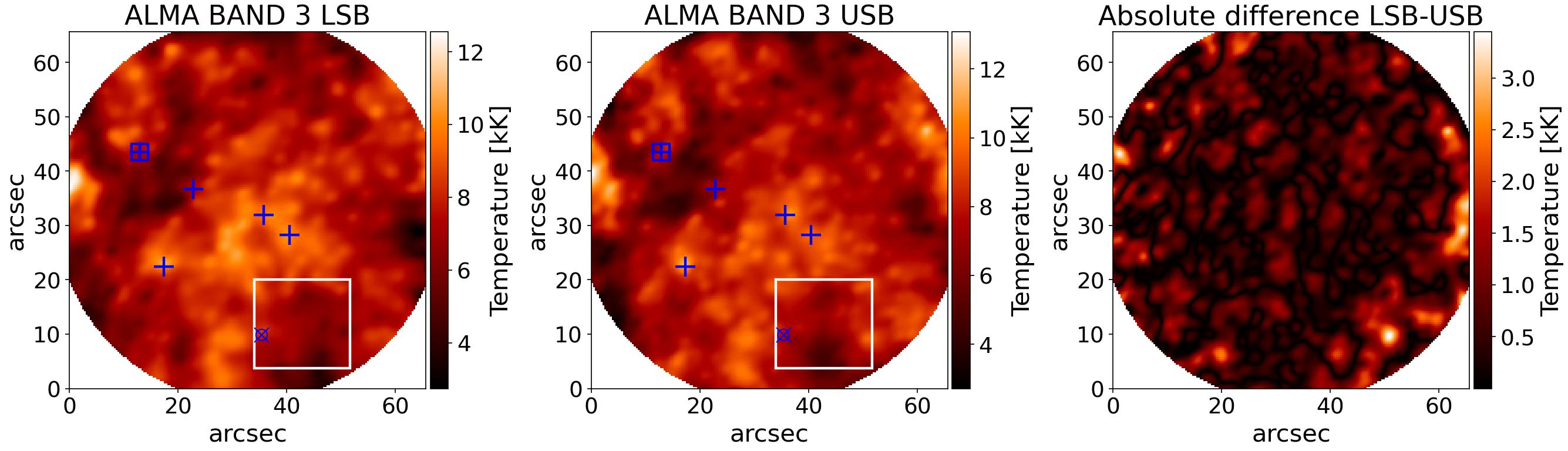}
    \caption{Same time frame for ALMA lower sideband LSB (left) and upper sideband USB (middle) and the absolute difference between the two sidebands (right). The white box depicts a relatively quiet region of the field of view. The blue crosses mark the location of five bright features analysed in here and the cross enclosed by the blue square marks the location of the feature whose transverse oscillation has been shown in Fig.~\ref{fig:feat_phase}. The blue crossed circle marks the pixel shown in Fig.~\ref{fig:phase_slope}.}
    \label{fig:ALMA_snaps}
\end{figure*}

Since  2016, ALMA has provided  high-quality, high-resolution observations of the Solar chromosphere at millimetre wavelengths, including the study of oscillations in Solar ALMA observations \citep{2020A&A...634A..86P,2020A&A...635A..71W, 2020A&A...644A.152E, 2021RSPTA.37900174J}. Thus far, however, Solar observations with ALMA have been limited to one receiver band at the time, only providing information about dynamical phenomena across  the (relatively small) height range from the radiation that emerges at the observed wavelengths. It should be noted that the absolute heights of formation vary from location and location as well as over time, as a result of the chromosphere's intermittent and dynamic nature \citep{2021A&A...656A..68E}. For Solar observations, each ALMA receiver band is organised into four sub-bands, with each individual sub-band consisting of several spectral channels spanning a certain range of frequencies. Observations at each channel take place in the same time, meaning that there is no time delay between measurements at different frequencies. Due to the particular challenges of observing the Sun with ALMA as compared to other targets, for instance, the antenna beam being filled with complex emission that varies on very short time scales, until very
recently, there was no standard reduction pipeline that could produce science ready time series of images. The standard approach so far is to use all the data to reconstruct one continuum map for each time step, thus neglecting the spectral domain in favour of higher image quality \citep{2022A&A...659A..31H}. In their studies, \citet{2019A&A...622A.150J} and \citet{2019ApJ...875..163R} split the data into the four sub-bands and analysed the resulting four sub-bands individually \citep[see also][]{2018A&A...617L...6R}. Since the first two sub-bands (i.e. SB-1 and SB-2) are directly adjacent in frequency, the corresponding SB-1 and SB-2 maps were found to be very similar. The same is true for the last two sub-bands (i.e. SB-3 and SB-4). However, close comparisons between individual sub-bands showed relatively low signal-to-noise ratios (S/N). Therefore, the Solar ALMA Pipeline (SoAP; Szydlarski et al., in prep.) was extended with an additional mode that reconstructs  image time series by using all data for SB-1 and SB-2 combined (together forming  the lower sideband, LSB) and respectively for SB-3 and SB-4 combined (together forming  the upper sideband, USB), which are simultaneous observations of the same target. The resulting two image time series for LSB (SB-12) and USB (SB-34) are found to have a higher S/N. 

In this letter, we exploit the new imaging mode and estimate the average formation height difference between the LSB/USB maps  by assuming that magnetically quiet regions are mostly dominated by purely acoustic waves. We further compare (in brief) the resulting height difference with that obtained from corresponding synthetic millimetre maps from magnetohydrodynamic simulations. Furthermore, we study transverse oscillations and we estimate the propagating speeds of the transverse wave and, ultimately, their average energy flux.

\section{ALMA sub-band observations}

The ALMA Band 3 (2.8-3.3\,mm) observation used in this study was carried out on 22 April 2017 between 17:20 and 17:55 UTC as part of program 2016.1.00050.S. This observation consists of a time series split into three scans (blocks of observation) with durations of about 10~min each and a cadence of 2~s. The spectral setup consists of four sub-bands with a bandwidth of 2\,GHz centred at different frequencies within the full-band range (SB1 is centred at 93\,GHz, SB2 at 95\,GHz, SB3 at 105\,GHz, and SB4 at 107\,GHz). For the purposes of this study, we combined the lower and upper pairs of sub-bands (within the reduction pipeline) to reconstruct the time series. Effectively, the first pair, hereafter referred to as LSB (SB-12), is a  4\,GHz band wide and centred at 94\,GHz; the second pair, hereafter referred to as USB (SB-34), is also 4\,GHz wide and centred at 106\,GHz. This approach allows to improve the S/N values compared to individual sub-bands. Moreover, it allows us to have co-temporal observations of the same region at two frequencies that are separated by a gap of 12\,GHz. 

The pixel size during the reconstruction of the time series was chosen to be 0.34\,arcsec for both LSB and USB. The spatial resolution is of about 2.1\,arcsec for the LSB and 1.9\,arcsec for the USB. The time series is reconstructed in such a way that the individual frames have the same size in LSB and USB, enabling  a pixel-to-pixel comparison between them. Furthermore, as the interferometric observation only provides relative differences in brightness temperature, the absolute temperature values were obtained by shifting the zero point  by 7418\,K in the case of LSB and 7277\,K in the case of USB, according to the average temperature values reported by  \citet{2022A&A...661L...4A}. The resulting brightness temperature ranges are [1974-13648]\,K, with a mean of 7418\,K and a standard deviation of 1417\,K for the LSB, and [1362-13946]\,K, with a mean of 7277\,K and a standard deviation of 1381\,K for the USB, respectively.

Figure~\ref{fig:ALMA_snaps} shows the same time frame for the LSB on the left and the USB on the middle, whereas the right panel shows the absolute temperature difference between the two sidebands. The latter clearly demonstrated that the LSB-USB differences provide valuable information regarding the thermal structure of the chromosphere. The ALMA maps are  spatially-coaligned with observations from the Solar Dynamic Observatory (SDO) \cite{2012SoPh..275....3P}. The Solar coordinates of the centres of the field of views (FOV) are $(x,y)\,=(-246,267)$\,(arcsec). The observation samples mainly a plage region on the east side of NOAA AR12651 but also a small, magnetically quiet region (marked with the white squares in the figure). A full description of the same observation although in the form of continuum-only (full-band) time series can be found in \citep{2021RSPTA.37900184G,2021RSPTA.37900174J,MHD_Guevara_Gomez}.

\section{Results} 

\subsection{Height differences}

\begin{figure}[tp!]
    \centering
    \includegraphics[width=0.45\textwidth]{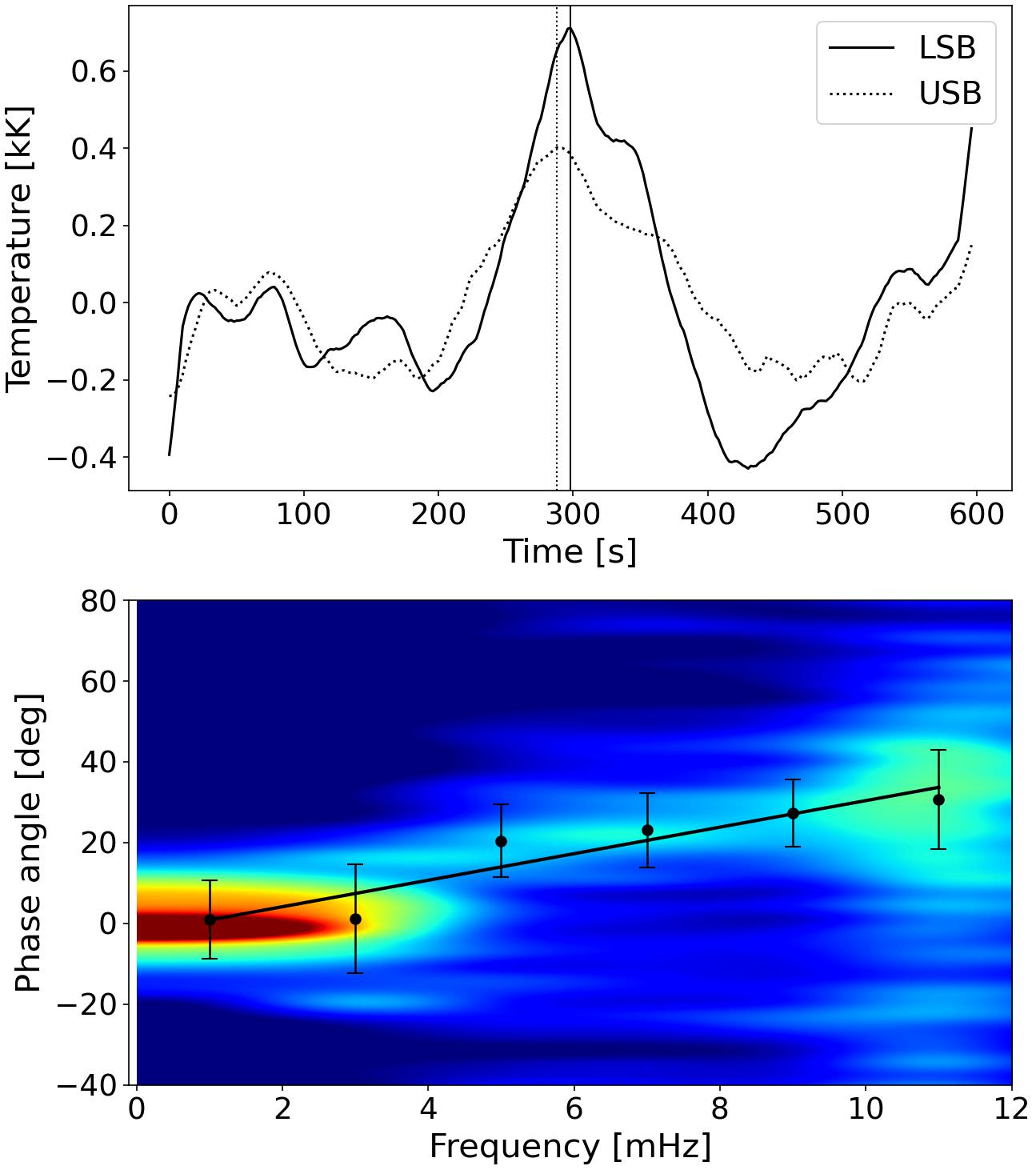}
    \caption{Brightness temperature delay between sidebands. \textit{Top:} Typical detrended brightness temperatures of the LSB (solid) and USB (dotted) for one pixel. The vertical solid and dotted lines mark the peak temperatures of the pixel for LSB and USB, respectively. \textit{Bottom:} Phase spectrum (2D histogram of phase-angle as a function of frequency) between ALMA Band 3 LSB and USB from the Fourier transform analysis of each individual pixel within the white box shown in Figure~\ref{fig:ALMA_snaps}. The slope, indicated with the black solid line, represents the wave travel time of acoustic waves. The error bars show 1$\sigma$ from Gaussian fitting.}
    \label{fig:phase_slope}
\end{figure}

The radiation at millimetre wavelengths is mainly formed via the interaction of electrons with the Coulomb field of charged ions, plus a relatively small contribution by neutral hydrogen affected by the Coulomb field of electrons passing by. These processes occur basically under local thermodynamic (LTE) conditions, so that the observed temperature at a certain frequency samples the predominant height over the Solar surface where the radiation at that frequency originates \citep{2016SSRv..200....1W,2020FrASS...7...57N}. In general, for a monotonic increase of temperature with height in the chromosphere, the formation height of the LSB is expected to be above the formation height of the USB as the centre frequency of the latter is 12\,GHz higher than the centre frequency of the former. Despite the fact that it is not possible to truly define what are the absolute formation heights at ALMA wavelengths based on these observations alone, \citet{2020A&A...640A..57A} estimated that the height difference between SDO/AIA 1600\,$\AA$ and ALMA Band 3 observations (centred at $100$\,GHz) is about 1200\,km, which would put the absolute formation height of ALMA Band 3 in the upper chromosphere. Furthermore, the study of oscillations in temperature at the two different sidebands LSB and USB can be used to estimate the relative height differences between them. For this purpose, we have chosen a $(17\times17)$\,(arcsec) or $(50\times50)$\,(pixels) box of a magnetically quiet area (see the white squares in Fig.~\ref{fig:ALMA_snaps}) within which acoustic waves with a speed of $c_s = 8.0\pm 1.0$\,km/s \citep[e.g.][]{2002ApJ...564..508R,2019AnGeo..37..891S} are expected to be dominant. Thus, we performed a Fourier analysis on the entire time series, individually for each pixel inside the white box, and we computed the phase-angles between the two sidebands from their cross spectra. A zero phase-angle between two signals means that the signals are in phase, while a non-zero phase-angle can imply that one signal is either leading or lagging behind the other one. Each phase-angle $\phi\,\text{[deg]}$ has a corresponding frequency $f\,\text{[Hz]}$ in the Fourier space, such that the two quantities are related with a time delay, $\tau\,\text{[s],}$ between the signals via the following equation:

\begin{equation}
    \phi\ = 360^{\circ} \,\tau\,f
    \label{eq:phase_freq}
.\end{equation}

The top panel of Fig.~\ref{fig:phase_slope} shows the detrended brightness temperatures for the LSB as a solid line and for the USB as a dotted line for the pixel marked with a crossed blue circle in Fig.~\ref{fig:ALMA_snaps}. This plot illustrates the typical behaviour of temperature within the white box corresponding to the magnetically quiet area. In the bottom panel of Fig.~\ref{fig:phase_slope}, the 2D phase spectrum of the corresponding temperature light curves between the LSB and the USB (for the quiet region) is shown. The black dots correspond to the centres of Gaussian curves fitted to vertical cuts in the spectrum and the error bars to their respective standard deviations. The solid black line corresponds to the linear regression fit to the black dots and shows the relation between phase-angle and frequency according to Eq.~\eqref{eq:phase_freq}. The slope of the line is given by $\Delta \phi / \Delta f = 360^{\circ} \,\tau$. Hence, the time delay between the ALMA LSB and USB is estimated to be  $\tau = 9.1 \pm 1.6$\,s. 

The time delay between LSB and USB is an indication of the travel time of a propagating wave observed in brightness-temperature oscillations. Under the assumption that the analysed magnetically quiet region is dominated by acoustic wave, it is then possible to obtain the (average) height difference $\Delta H$ between the two ALMA sidebands as $\Delta H = c_s\tau$. Using the values derived above, we obtain $\Delta H \approx 73 \pm 16$\,km. This value falls within the range of formation height differences between synthetic continuum maps calculated for the public enhanced network Bifrost simulation \citep{2016A&A...585A...4C} for  the same ALMA sidebands. Specifically, the height differences in the simulation for a relatively quiet region are predominantly distributed between 20\,km and 120\,km with the peak of the distribution at $\approx$60\,km.

We note that based on the theory presented by \cite{2006ApJ...640.1153C}, \citet{2017ApJS..229...10J} used an identical method to estimate the height difference between the 300\,nm and the Ca\,{\sc{ii}}\,H 396.8\,nm passbands of the filter imager on board the {\sc{Sunrise}} balloon-borne solar observatory \citep{2010ApJ...723L.127S}. As such, we refer to \citet{2017ApJS..229...10J} for further details on this approach.

\subsection{Propagation of transverse waves}

A statistical study of the same ALMA observations (prepared, instead, as full-band maps) showed the possible presence of MHD transverse (kink) oscillations in small-scale bright features. Specifically, \citet{MHD_Guevara_Gomez} analysed $\approx$ 200 bright features in ALMA Band 3 (full-band), which exhibited transverse oscillations in the horizontal velocities. The amplitude of the oscillations spanned a range between 0.2-27.1\,km/s with an average oscillation period of 66\,s. These properties suggested that the transverse oscillations may be associated with kink MHD modes \citep[see e.g.][and references therein]{2015SSRv..190..103J}. In this letter, we have selected five magnetic bright features to study their transverse-oscillatory properties in the two sidebands. Their median locations are marked with blue crosses in  Fig.~\ref{fig:ALMA_snaps}. Each of the features is visible and traceable in time in the two sidebands. The border of the features is defined by the contour at half of the maximum temperature of the features at each frame. The location of the features is computed as the centre of gravity (of intensity) using the temperatures within the feature borders. For each feature, the total horizontal velocity is calculated as $\mathrm{v_t} = \sqrt{\mathrm{v_x}^2 + \mathrm{v_y}^2,}$ where the velocities in $x$ and $y$ directions correspond to the displacement of the centre of  gravity from frame to frame in each direction. The total horizontal velocities show a similar behaviour as those analysed in \cite{2021RSPTA.37900184G,MHD_Guevara_Gomez} suggesting the presence of MHD kink modes.

\begin{figure}[htp!]
    \centering
    \includegraphics[width=0.45\textwidth]{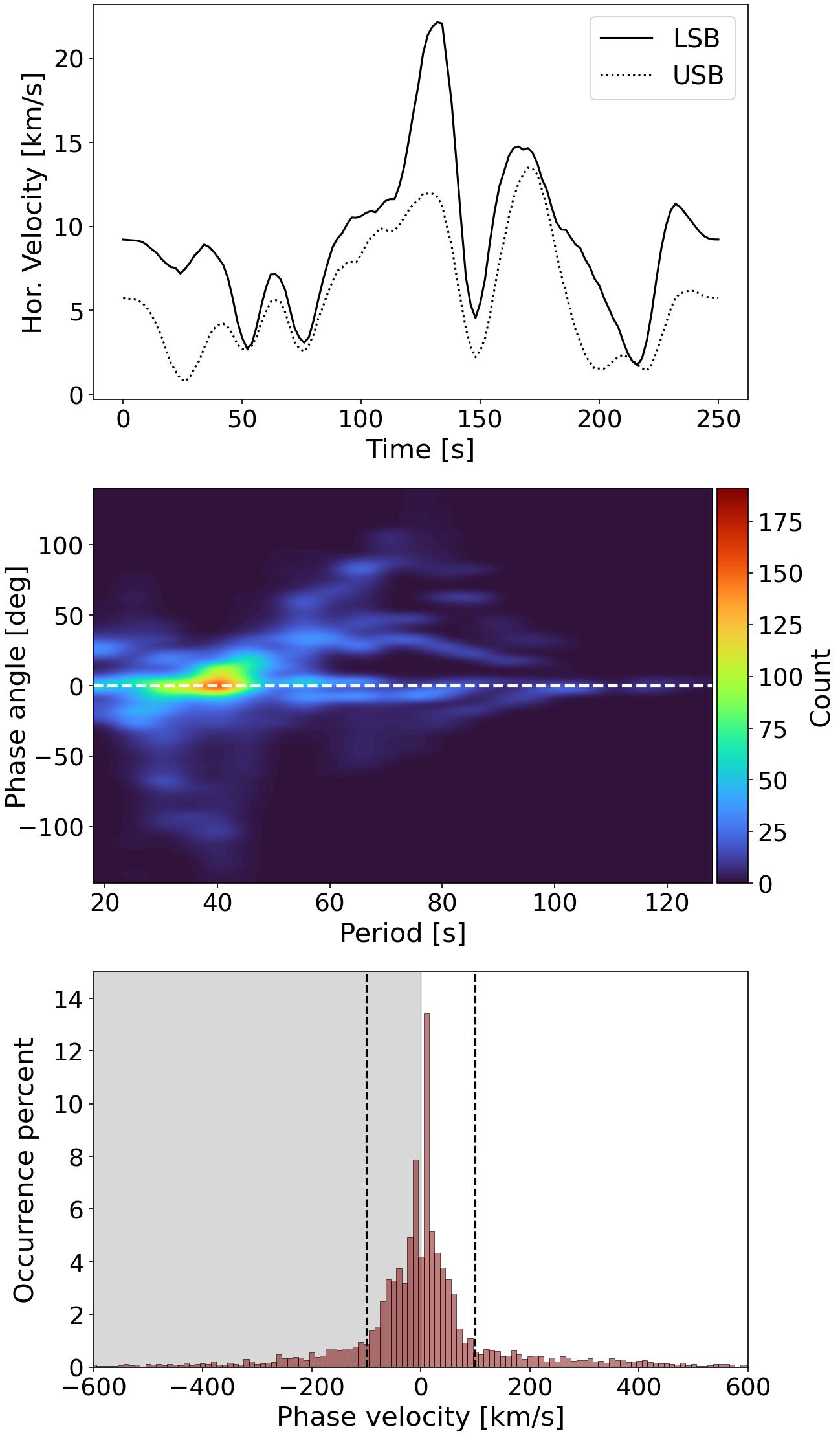}
    \caption{Propagation of transverse waves between sidebands. \textit{Top:}~ Horizontal velocity for the bright feature enclosed in a square in Fig.~\ref{fig:ALMA_snaps}. \textit{Middle:}~Phase diagram as a 2D histogram of phase angle vs periods of the horizontal velocity oscillations in the five small bright features (see blue crosses in Fig.~\ref{fig:ALMA_snaps}) observed simultaneously in the LSB and the USB. Positive and negative phase angles represent upward and downward propagation in the Solar chromosphere, respectively. \textit{Bottom:}~Histogram showing the calculated phase velocities for the waves present in the five analysed features, with a bin width of 10\,km/s. }
    \label{fig:feat_phase}
\end{figure}

For each individual feature, we have computed the phase-angles between velocity oscillations observed in the two sidebands by the means of a cross-wavelet transform analysis. Each phase-angle is associated to a dominant period of oscillation within the 95\% confidence level used in the wavelet (i.e. regions on the wavelet spectra where the power exceeds a 95\% confidence level and is outside the cone of influence). By putting together all the phase-period values identified in the five features, it is possible to draw a phase diagram of the horizontal displacements. We note that there are several phase angles associated to each bright feature. Figure~\ref{fig:feat_phase} shows in the top a typical transverse oscillation of a feature as a plot of the horizontal velocity versus time for the feature marked with a blue cross enclosed by a square in Fig.~\ref{fig:ALMA_snaps}. In the middle of the figure, we show the phase diagram in the form of a 2D histogram. The brightest part close to a zero phase and a period of 40\,s indicates the maximum occurrence of standing waves, namely, where there is no propagation. Then, the slightly less strong occurrence above and below the white dotted line would be due to upwardly and downward propagating waves, respectively.

Although the absolute formation heights vary from location to location and they are shown to be higher in magnetic elements compared to quiet regions \citep{2015A&A...575A..15L, MHD_Guevara_Gomez}, their average height difference should not change considerably (see e.g. Figure~2 of \citet{2017ApJS..229...10J}, where the average height difference for quiet-Sun and plage models are similar). Therefore, it is practical to use the value of $\Delta H \approx 73 \pm 16$\,km that was previously found to estimate the velocity propagation (phase velocity $\mathrm{v_{ph}}$) of the transverse waves through the chromospheric layers probed with ALMA sidebands. To this end, we use the following equation for the phase velocity 

\begin{equation}
    \mathrm{v_{ph}}\,=\,\dfrac{360^{\circ}\Delta H}{T\varphi}
    \label{eq:vel}
,\end{equation}

where $T$ is the period in seconds and $\varphi$ is the phase-angle in degrees. The distribution of the resulting phase velocity values is presented in the bottom panel of Fig.~\ref{fig:feat_phase}, where negative values correspond to downward propagation and positive values to upward propagation. The time lags corresponding to phases of 0$^{\circ}$ are interpreted as standing waves between the two layers and they are therefore excluded of the distribution. However, they represent about $6\%$ of the  total number of occurrences. This procedure is similar to the method used by \citet{2022ApJ...930..129B} to estimate the phase velocities of transverse oscillations in spicules. The distribution of velocities shows that more than $72\%$ of the computed values are between -100 and 100\,km/s. The occurrence rate of non-standing waves were 54\% and 46\% for those corresponding to upwardly and downwardly propagation, respectively. The mean and median upward phase velocities are 94\,km/s and 36\,km/s. For the downward phase velocities the mean and the median are $-99$\,km/s and $-53$\,km/s, whereas the absolute mean velocity is 96\,km/s. 

Under the assumption that the horizontal oscillations are related to MHD kink waves, it is possible to estimate the energy flux $F$ that the identified kink waves carry between the two heights according to \citet{2015A&A...578A..60M} with the following equation:

\begin{equation}
    F\,\approx \dfrac{1}{2}\,f\,(1+\ln{1/f})\,\rho\,\mathrm{v_{amp}^2}\,\mathrm{v_{ph}}
    \label{eq:Energy}
,\end{equation}

where $\rho = 2.33\times 10^{-8}$\,kg/m$^3$ is the density just outside the waveguide, $f=0.045$ is a filling factor, and $\mathrm{v_{amp}}=4.3$\,km/s is the amplitude of the velocity oscillations. These quantities were taken from the statistical values reported by \citet{MHD_Guevara_Gomez}. We take $\mathrm{v_{ph}}=96$\,km/s as the absolute mean velocity and replace the values in Eq.~\eqref{eq:Energy}, obtaining an energy flux of about $3800$\,W/m$^2$.

\section{Conclusions}

We first studied the simultaneously observed temperature oscillations in a magnetically quiet region observed at two different frequencies with ALMA. To achieve this, we made used of a special procedure to reconstruct time series of two sidebands within the ALMA Band 3 receiver, that is, LSB (94\,GHz) and USB (106\,GH).
Under the assumption that the temperature fluctuations in the relatively quiet region of the LSB and USB represent propagating acoustic waves with a speed of $8 \pm 1$\,km/s, we computed a height difference of $73 \pm 16$\,km between the two chromospheric layers from the phase differences between the temperatures oscillations in LSB and USB maps. This study demonstrates the potential diagnostic use of ALMA LSB and USB observations to probe the Solar atmosphere.

Furthermore, we traced and studied the oscillatory properties of five small-scale bright magnetic features present in both sidebands. In particular, we compared the transverse oscillation of the features through wavelet analysis, resulting in the detection of  positive and negative phase lags between them, namely, upwardly and downwardly propagating waves, with mean velocities of $94$\,km/s and $-99$\,km/s, respectively. These phase velocities are comparable to those of fast kink waves observed in spicules and fibrils (in the Solar chromosphere) with velocities on the order of 50-150\,km/s \citep{2009A&A...497..525H,2011ApJ...736L..24O,2012NatCo...3.1315M,2015SSRv..190..103J,2017ApJS..229....9J}. Taking into account that these velocities correspond to waves propagating upwardly and downwardly in the chromosphere, using a vertical or near-vertical magnetic field as wave guides (see the magnetic topology of the same data set in \citealt{2021RSPTA.37900174J}), we speculate that the observed features may be spicules seen from the top. The presence of standing waves with a strong occurrence over periods close to 40\,s was identified as well. The standing waves may be due to a superposition of upward and downward propagating waves, the latter being the product of reflections of the former somewhere near to the transition region, above the heights mapped with ALMA.

Finally, we also estimated the energy flux carried  by the propagating kink waves to be on average about $3.8\times 10^3$\,W/m$^2$, which is close to the value of $4\times 10^3$\,W/m$^2$ needed to compensate for radiative losses in the chromosphere according to \citet{1977ARA&A..15..363W}. However, this does not imply that the energy carried by these waves is completely dissipated in the chromosphere and therefore able to account for the radiative losses alone; it indicates, rather, that their contribution to sustain a hot chromosphere may be substantial. An in-depth analysis of these waves, as well as other MHD modes, from both ALMA sidebands observations and numerical simulations, is essential for identifying how they can contribute to  Solar atmospheric heating and this will be the subject of a future work.

\section*{Acknowledgments}

This work is supported by the SolarALMA project, which received funding from the European Research Council (ERC) under the European Union’s Horizon 2020 research and innovation programme (grant agreement No. 682462), and by the Research Council of Norway through its Centres of Excellence scheme, project number 262622. 
This paper makes use of the following ALMA data: ADS/JAO.ALMA\#2016.1.00050.S. ALMA is a partnership of ESO (representing its member states), NSF (USA) and NINS (Japan), together with NRC(Canada), MOST and ASIAA (Taiwan), and KASI (Republic of Korea), in co-operation with the Republic of Chile. The Joint ALMA Observatory is operated by ESO, AUI/NRAO and NAOJ. We are grateful to the many colleagues who contributed to developing the Solar observing modes for ALMA and for support from the ALMA regional centres.

\bibliographystyle{aa}
\bibliography{ref.bib}

\end{document}